\documentclass[pra,onecolumn,showpacs,superscriptaddress,nofootinbib]{revtex4-2}
\usepackage{blindtext}
\usepackage{graphicx}
\usepackage{booktabs}
\usepackage{tabularx}
\usepackage{multirow}

\usepackage[table, svgnames, dvipsnames]{xcolor}
\usepackage{makecell, cellspace}
\usepackage{bbold}
\usepackage{amsmath, dsfont}
\usepackage{amsfonts}
\usepackage{bbm}
\usepackage{mathdesign}
\usepackage{mathtools}
\usepackage{amsthm}
\usepackage{bm}
\usepackage{amssymb}
\usepackage[normalem]{ulem}
\usepackage{braket}
\usepackage{hyperref}
\hypersetup{colorlinks=true,allcolors=black}
\usepackage{tikz}
\usepackage{ifthen}
\usepackage{stmaryrd}
\usetikzlibrary{tikzmark}
\usepackage{upgreek}
\usepackage{physics}
\usepackage{placeins}
\makeatletter
\let\origFloatBarrier\FloatBarrier
\renewcommand{\FloatBarrier}{\par\origFloatBarrier}
\makeatother
\usepackage{ragged2e} 

\usepackage[
    top=3cm,    
    bottom=3cm, 
    left=2cm, 
    right=2cm 
]{geometry}

\usepackage{svg}

\usepackage{booktabs, siunitx}
\usepackage{ragged2e} 

\usepackage{graphicx}
\usepackage[shortlabels]{enumitem}

\usepackage{tikz-cd}
\usepackage{systeme}
\usepackage{bm}
\usepackage{tensor}
\usepackage{xparse}
\usepackage{amsmath}
\usepackage{physics}
\usepackage{amsthm}
\usepackage{mathtools}
\usepackage{amssymb}
\usepackage{dsfont}
\usepackage{xcolor}
\usepackage{upgreek}
\usepackage{mathrsfs}

\definecolor{mycolor}{HTML}{90E0EF}
\usepackage{mathrsfs}
\usepackage{hyperref}

\usepackage{lettrine}
\usepackage{setspace}
\usepackage[utf8]{inputenc} 
\usepackage[T1]{fontenc}

\usepackage{wrapfig}

\renewcommand{\selectlanguage}[1]{}

\definecolor{blue}{HTML}{03045E}
\definecolor{red}{HTML}{c1121f}

\usepackage{orcidlink}
\hypersetup{
	colorlinks=true,
	citecolor= blue,
	linkcolor=blue,
	filecolor=blue,      
	urlcolor=blue,
	pdftitle={Overleaf Example},
	pdfpagemode=FullScreen,
}

\usepackage[T1]{fontenc}
\usepackage[utf8]{inputenc}

\usepackage{titlesec}

\titleformat{\section}{\sffamily\bfseries}{\thesection}{1em}{}
\titleformat{\subsection}{\sffamily\bfseries}{\thesubsection}{1em}{}

\titleformat{\subsubsection}
  {\normalsize\itshape\centering} 
  {\thesubsubsection.} 
  {1em} 
  {} 

\titlespacing\section{0pt}{12pt plus 3pt minus 2pt}{5pt plus 2pt minus 2pt}
\titlespacing\subsection{0pt}{12pt plus 3pt minus 2pt}{5pt plus 2pt minus 2pt}
\titlespacing\subsubsection{0pt}{12pt plus 3pt minus 2pt}{5pt plus 2pt minus 2pt}

\usepackage{floatrow}
\usepackage{graphicx}
\usepackage[label font=bf,labelformat=simple]{subfig}
\usepackage{caption}

\usepackage{hyperref}
\UseRawInputEncoding
\definecolor{blue}{HTML}{03045E}
\definecolor{violet}{HTML}{8338EC}
\definecolor{yale_blue}{HTML}{40916C}

\usepackage[normalem]{ulem}

\usepackage{physics}
\DeclareMathOperator*{\argmin}{argmin}
\usepackage{verbatim}

\newcounter{scheme}
\renewcommand{\thescheme}{\arabic{scheme}}

\usepackage{bibunits}
\defaultbibliographystyle{apsrev4-2}
\defaultbibliography{biblio}

\begin{document}
\fontsize{10.5}{12.6}\selectfont
\color{black}
\begin{bibunit}

\title{\textbf{Accelerating \textit{De Novo} Genome Assembly via Quantum-Assisted Graph Optimization with Bitstring Recovery}}

\author{Jaya Vasavi Pamidimukkala}
\thanks{These authors contributed equally to as first authors.}
\affiliation{Department of Biotechnology, Indian Institute of Technology, Madras, Chennai, India}

\author{Himanshu Sahu}
\thanks{These authors contributed equally to as first authors.}

\affiliation{Department of Physics and Department of Instrumentation \& Applied Physics, Indian Institute of Sciences, Bangalore, India.}

\author{Ashwini Kannan}
\thanks{These authors contributed equally to as second authors.}
\affiliation{Department of Biotechnology, Indian Institute of Technology, Madras, Chennai, India}

\author{Janani Ananthanarayanan}
\thanks{These authors contributed equally to as second authors.}
\affiliation{IBM Research, Bangalore, India}

\author{Kalyan Dasgupta}\email{kalyand1@in.ibm.com}

\affiliation{IBM Research, Bangalore, India}

\author{Sanjib Senapati}\email{sanjibs@iitm.ac.in}
\affiliation{Department of Biotechnology, Indian Institute of Technology, Madras, Chennai, India}

\begin{abstract}
\normalsize
\textsf{\textbf{Abstract}:} Genome sequencing is essential to decode genetic information, identify organisms, understand diseases and advance personalized medicine. The sequencing process involves fragmenting the entire genome, identifying each fragment experimentally, and then computationally assembling them to reconstruct the genome. A critical challenge lies in assembling the small genome fragments into the correct order. However, the current \textit{de novo} genome assembly processes, which involve constructing an entire genome sequence from scratch without a reference genome, present significant challenges due to its high computational complexity, affecting both time and accuracy. In this study, we propose a hybrid approach utilizing a quantum computing-based optimization algorithm integrated with classical pre-processing to expedite the genome assembly process. Specifically, we present a method to solve the Hamiltonian and Eulerian paths within the genome assembly graph using gate-based quantum computing through a Higher-Order Binary Optimization (HOBO) formulation with the Variational Quantum Eigensolver algorithm (VQE), in addition to a novel bitstring recovery mechanism to improve optimizer traversal of the solution space. The HOBO formulation reduces the qubit requirement to $\mathcal{O}(N\log_2 N)$. A comparative analysis with classical optimization techniques was performed to assess the effectiveness of our quantum-based approach in genome assembly. The experiments spanned up to systems having $24$ nodes involving $120$ qubits in IBM hardware. The results indicate that, as quantum hardware continues to evolve and noise levels diminish, our formulation holds a significant potential to accelerate genome sequencing by offering faster and more accurate solutions to the complex challenges in genomic research.
\normalsize
\end{abstract}

\maketitle

\section{Introduction}
Whole genome sequencing is a revolutionary progress in biomedical research that involves sequencing the entire genome of an organism that can aid in annotating new organisms, exploring evolutionary relationships, studying the genetic makeup of pathogens, identifying carcinogenic mutations, advancing forensic investigations, and developing personalized medicine. Genome sequencing works by fragmenting the entire genome, identifying each fragment experimentally, and computationally assembling them to reconstruct the genome. A critical challenge arises in\textit{ de novo} genome sequencing, where no reference genome is available, requiring accurate assembly of small genome fragments in the correct order.

\textit{De novo} genome assembly complexity scales non-linearly with genome size. Small bacterial genomes of approximately $5$--$10\times 10^3$ bases can be assembled within a few hours whereas assembling the $3.2\times 10^6$-base-long human genome demands high-performance computing systems, with memory requirements reaching 646~GB and more than $150{,}000$ CPU hours using conventional assembly pipelines~\cite{digenova2021wengan}. As genome assembly is computationally expensive, particularly for large and repetitive genomes, quantum computing has been proposed as a potential acceleration strategy. Quantum superposition could enable the simultaneous exploration of multiple sequence alignments and assembly paths, thereby improving the efficiency of large-scale \textit{de novo} assembly processes.

Various genome sequencing technologies, such as Illumina~\cite{BUERMANS20141932}, Nanopore~\cite{Wang2021_NanoporeReview}, and real-time sequencing platforms, generate short or long reads that are assembled into complete genome sequences. For short-read data, genome assembly commonly relies on the de Bruijn graph (DBG) approach, where reads are divided into $k$-mers and assembled by identifying Eulerian paths or employing other efficient graph traversal methods. This approach is both accurate and computationally effective. In contrast, long-read sequencing data are typically assembled using the Overlap-Layout-Consensus (OLC) strategy, which constructs an overlap graph and solves it via a Hamiltonian path formulation. Most common DBG-based assemblers are \textit{SPAdes}~\cite{Bankevich2012_SPAdes}, \textit{ABySS}~\cite{Simpson2009_ABySS}, and \textit{Velvet}~\cite{Zerbino2008_Velvet}, while popular OLC assemblers include \textit{Canu}~\cite{Koren2017_Canu}, \textit{Celera Assembler}~\cite{Denisov2008_Celera}, and \textit{FALCON}~\cite{Chin2016_FALCON}. The underlying computational complexity differs significantly between these approaches: the Eulerian path problem is polynomial and hence tractable, whereas the Hamiltonian path problem is NP-complete. The runtime of these classical assemblers depends on multiple factors such as genome size, sequence complexity, and data quality. Typically, assembling small genomes may take from a few hours to a full day. Algorithmically, DBG assemblers exhibit a computational complexity of $O(N+E)$ for both Eulerian path and general graph traversal methods, where $N$ and $E$ denote the number of nodes and edges, respectively ~\cite{compeau_how_2011}. In contrast, solving the DBG or OLC assembly using a Hamiltonian path approach has a factorial complexity of $O(N \times N!)$, making it computationally expensive for large datasets \cite{gfg_hamiltonian_dp}.

Although solving the Hamiltonian path is an NP-complete problem, long-read sequencing provides a powerful approach for detecting genetic variants and resolving repetitive regions that are challenging to capture with short-read methods. Short-read sequencers generate reads up to 600 bases, offering high accuracy, low cost, and a wide array of analysis tools. Whereas, long-read technologies produce much longer reads, often exceeding 10 kb, which facilitates tasks such as \textit{de novo} genome assembly, accurate mapping, detection of structural variations, and identification of transcript isoforms \cite{pmc7006217}. By sequencing native DNA or RNA directly, long-read approaches avoid amplification bias and preserve important base modifications. With ongoing improvements in accuracy, throughput, and cost, long-read sequencing is increasingly accessible and reliable for genomic studies in both well-characterized and poorly studied species. Given the computational complexity associated with the Hamiltonian Path problem and its relevance to long-read assembly, recent studies are exploring quantum computing approaches as a means to address this problem more efficiently.

Well-established quadratic formulations exist for solving the Hamiltonian path problem, and a growing body of literature explores the use of quantum computers for this task. These approaches include both gate-based methods and quantum annealing techniques. We first review gate-based approaches.

Sarkar A et al. \cite{Sarkar} introduced QuASeR (Quantum Accelerated de novo DNA Sequence Reconstruction), a framework for addressing the\textit{ de novo }genome assembly problem. Their approach employs a QUBO (Quadratic Unconstrained Binary
Optimization) formulation of the travelling salesman problem, which is equivalent to the Hamiltonian path problem. The formulation is implemented using both annealing-based simulators and D-Wave quantum annealers, as well as gate-based quantum simulators. For the gate-based implementation, the authors apply the Quantum Approximate Optimization Algorithm (QAOA) to identify the optimal path. Their experiments demonstrate results for a four-node system with reads consisting of ten nucleotide bases. Similarly, Fang et al. \cite{Fang} proposed a divide-and-conquer algorithm for \textit{de novo} genome assembly using hybrid long and short reads. In this approach, the Eulerian path is computed on DBGs constructed from short reads, while long reads provide long-range constraints. The Eulerian path problem is solved using a Variational Quantum Eigensolver (VQE) implemented on a gate-based quantum simulator. Constraints are imposed to ensure path continuity and prevent branching in the assembly. A one-hot encoding scheme is employed, which is closely related to a QUBO formulation. The authors also investigate different ansatz choices for solving the VQE problem within the simulator. 

In \cite{Varsamis}, the authors propose a method for \textit{de novo} DNA sequence assembly using quantum random walks on graphs represented as Ising Hamiltonians. Their approach recursively partitions the graphs into smaller subgraphs using a max-cut algorithm, after which quantum walk algorithms are applied to identify paths in graphs with low hierarchical rank. The max-cut problem is solved using the QAOA algorithm in Qiskit. In a related but distinct line of work, the authors in \cite{kosoglu-kind_biological_2023} a biological sequence comparison algorithm using quantum computers. They utilize the Flexible Representation of Quantum Images (FRQI) framework for comparisons at a granular level. The position of the base and its value are encoded to a quantum state. 

A substantial body of work in the literature also explores annealing-based approaches. Anuradha et al. \cite{Mahasinghe} present a QUBO formulation tailored for adiabatic quantum computation. The authors analyze the complexity of implementing this formulation on a D-Wave quantum computer and, prior to introducing their approach, compare it with existing methods based on Grover's algorithm and quantum walks. In a related study, Boev et al. \cite{boev_genome_2021} introduce an annealing-based approach for \textit{de novo} genome assembly, demonstrating their method on both simulated data and the $\phi X174$ bacteriophage. Their work employs a well-known QUBO formulation and is implemented on a D-Wave quantum computer. A key challenge addressed in this study is mapping the native connectivity of the physical qubits onto the problem Hamiltonian.

As highlighted by the literature, most existing algorithms rely on QUBO formulations, in which the position of each node in the graph is represented by one-hot encoding of binary variables. Such formulations typically require a large number of qubits for encoding, limiting their scalability. In this work, we propose a Higher-Order Binary Optimization (HOBO) approach, which reduces the qubit requirement to $N \log_2 N$, where $N$ is the number of nodes in the graph. Using this framework, we demonstrate quantum-assisted OLC graph solving on a subset of the genome graph, enabling efficient reconstruction of partial genome sequences. This approach not only facilitates accurate organism identification but also offers improved computational efficiency and faster execution. We also employ a novel bitstring recovery mechanism to handle the problem specific errors we see arising in results when the problem is scaled up to a higher number of nodes. 

Our paper is organized as follows. The detailed end-to-end methodology (read preprocessing, graph construction, classical dynamic programming benchmark, variational quantum workflow, bitstring recovery, and contig generation) is provided in the Supplementary Methodology (Supplementary File, Section~\ref{supp:method}). In the main text, we focus on the HOBO formulation, the associated cost function and its complexity, approaches to designing the ansatz, and results from our runs in quantum simulators and IBM quantum hardware, followed by conclusions.

\section{Methodology}
This section outlines the workflow of the hybrid quantum-classical methodology for assembling sequencing reads, with detailed procedures described in the Supplementary Methodology (Supplementary File, Section~\ref{supp:method}). Genome assembly graphs are generated from sequencing reads, solved to determine optimal paths, and used to construct contigs representing partial genome sequences. We also outline the quantum algorithm used for graph resolution and the bitstring recovery mechanism designed to improve optimizer traversal of the solution space. An overview of the workflow is shown in Scheme~\ref{sch:workflow}.

\par
{\begingroup
\stepcounter{scheme}
\renewcommand{\thefigure}{\thescheme}
\makeatletter
\renewcommand{\fnum@figure}{Scheme~\thescheme}
\makeatother
\begin{figure}
	\centering
	\includegraphics[width=0.8\linewidth]{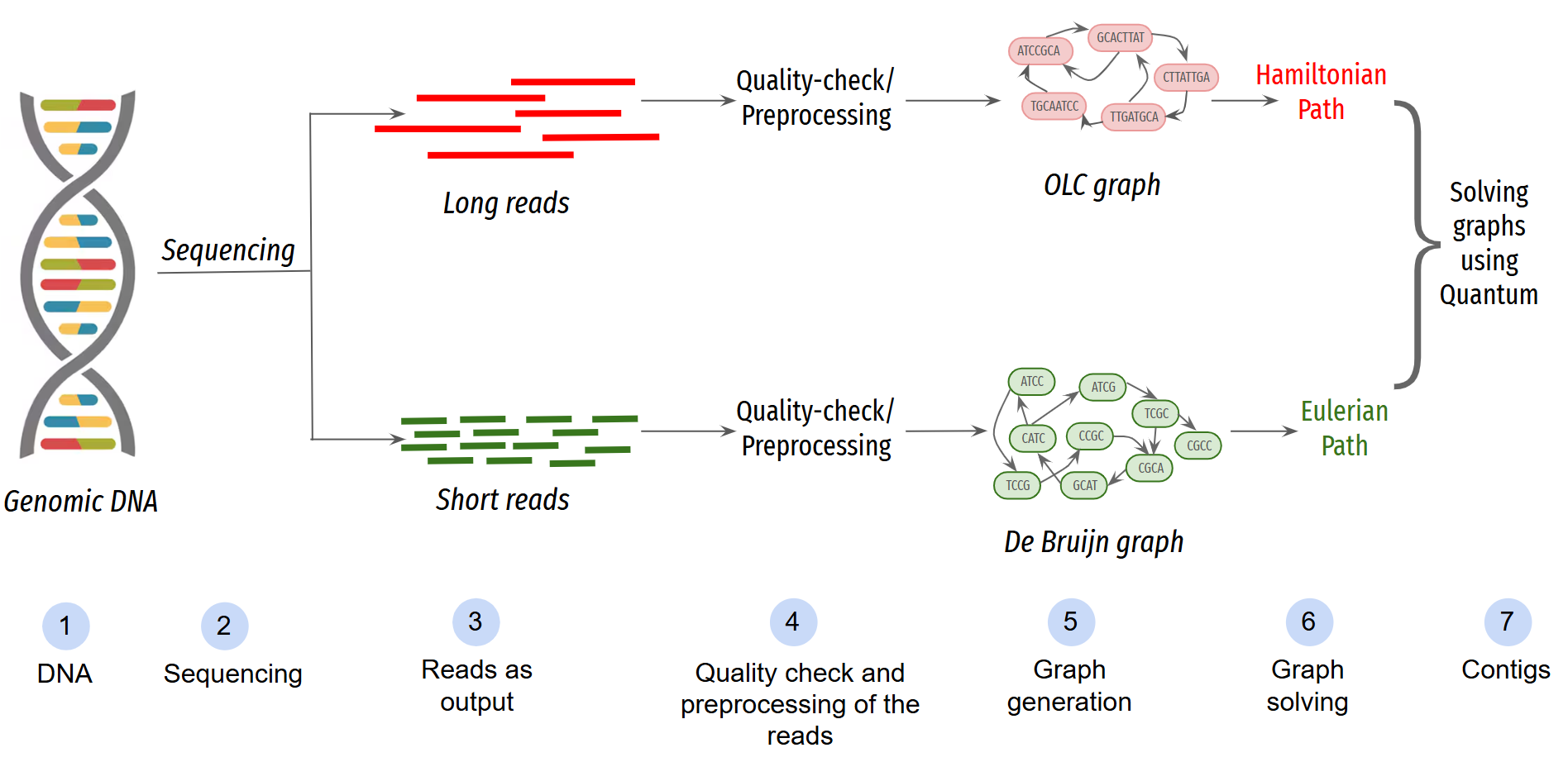}
	\caption{A streamlined pipeline for efficient genome sequencing using hybrid classical-quantum methods.}
	\label{sch:workflow}
\end{figure}
\endgroup}
\addtocounter{figure}{-1}
\FloatBarrier

\section{Proposed HOBO Reformulation to Solve Genome Assembly Problem} \label{sec:HOBO}
Variational quantum algorithms (VQAs)\,\cite{RevModPhys.94.015004,cerezo_variational_2021} are promising candidate hybrid algorithms for observing quantum computation utility on noisy near-term devices. Variational algorithms are characterized by the use of a classical optimization algorithm to iteratively update a parameterized trial solution, or "ansatz". Chief among these methods is the Variational Quantum Eigensolver (VQE), which aims to solve for the ground state of the given Hamiltonian represented as a linear combination of Pauli terms, with an ansatz circuit where the number of parameters to optimize over is polynomial in the number of qubits. Executing a VQE algorithm requires the following components: Hamiltonian and ansatz (problem-specific), estimator, and classical optimizer (see Supplementary Section~\ref{method:VQA}).

\subsection{Cost Function}
Consider the directed graph $\mathcal{G} = (\mathcal{V},\mathcal{E})$ such that $|\mathcal{V}|=N$ representing the number of nodes. In case of OLC algorithm, the node $i\in \mathcal{V}$ represents a sequencing read, and the goal is to find a Hamiltonian path in the graph. Each edge $e = (i,j)\in \mathcal{E}$ has a associated weight $W_{ij}$ (cost matrix) determining the overalp between the reads which are set to a value $-\alpha$ if associated otherwise to a large value $\gamma$ as a penalty. To formulate the cost function, define $b_{ti}$ to be a binary variable such that $b_{ti} = 1$ if the $i$-th node is placed at position $t$ ($t$ could be a time instance or an instance of a spatial position). The cost function takes the form
\begin{align}
H(b) = {}& A_1 \sum_{t=0}^{N-1} H_\text{valid}(b_t)
\\
&+ A_2 \sum_{t=0}^{N-1} \sum_{t'=t+1}^{N-1} H_{\not =}(b_t,b_{t'})
\\
&+ \sum_{i,j=0}^{N-1} W_{ij} \sum_{t=0}^{N-1} b_{ti} b_{t+1,j}\, \label{eq:cost-function}
\end{align}

\begin{figure*}[ht]
	\centering
	\includegraphics[width=0.85\linewidth]{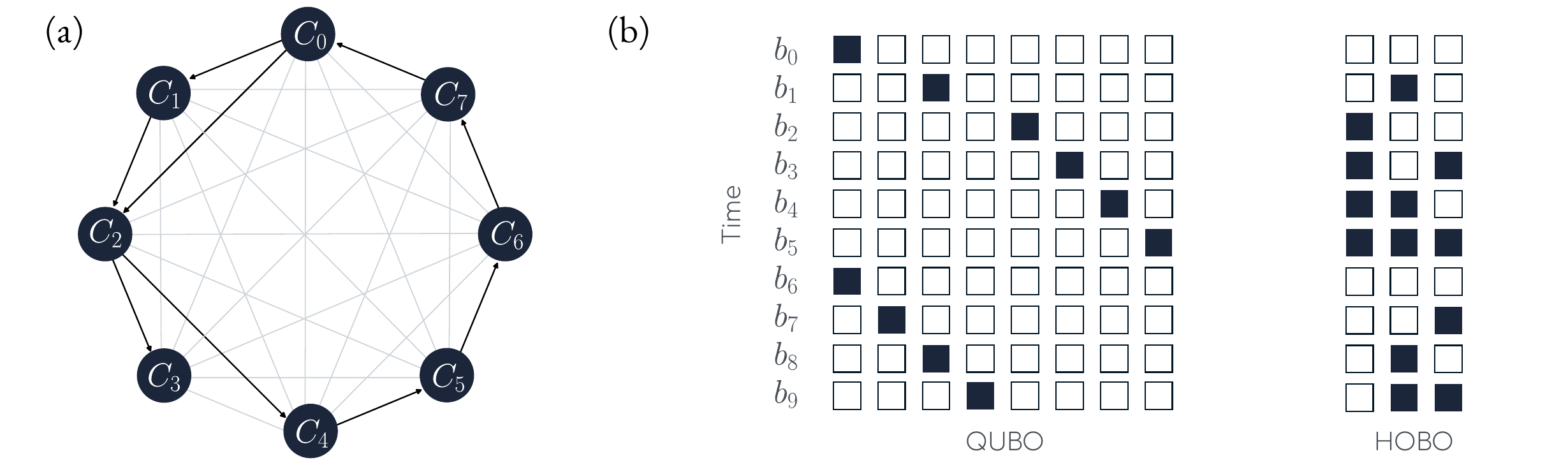}
	\caption{Visualization of qubit encodings for the genome sequencing problem. (a) An example graph where black edges represents the connected nodes, while gray edges shows unconnected nodes. (b) QUBO encoding (c) HOBO encoding}
	\label{fig:encoding}
\end{figure*}

Here $A_1,A_2 > \max W_{ij}$ are parameters that have to be adjusted during the optimization. In Eq.~\eqref{eq:cost-function}, $H_\text{valid}$ ensures that a vector of bits $b_t$ encodes a valid node at all instances of time/position, and $H_{\not =}$ ensures no two instances of time/position encode the same node. In QUBO encoding, each node is represented by an one-hot vector, see Fig.~\ref{fig:encoding}. The cost function in QUBO encoding takes the form 
\begin{align}
    & H^{\text{QUBO}}(b) = A_1 \sum_{t=0}^{N-1}\left(1- \sum_{i=0}^{N-1} b_{ti}\right) + \nonumber  \\ 
    & A_2 \sum_{i=0}^{N-1} \left(1- \sum_{t=0}^{N-1} b_{ti}\right) + \sum_{i,j=0}^{N-1}W_{ij}\sum_{t=0}^{N-1} b_{ti}b_{t+1,j} \,\label{eq:qubo}
\end{align}
The possible assembly path can be expressed as a quantum state 
\begin{equation}
	\begin{split}
|\Psi\rangle &= \bigotimes_{t=0}^{N-1} |b_{0t}b_{1t}\cdots b_{N-1,t}\rangle  \,
\end{split}
\end{equation}
The standard technique in QUBO formulation\,\cite{lucas_ising_2014} is the substitution
\begin{equation}b_{i,t} \rightarrow  I^{(0)} \otimes \cdots \otimes \mathcal{Q}^{(t)}_i \otimes \cdots \otimes I^{(t_\text{max}-1)} \end{equation}
with 
\begin{equation} \mathcal{Q}^{(t)}_i = I_0 \otimes \cdots \otimes I_{i-1}\otimes   P_- \otimes I_{i+1} \otimes \cdots\otimes I_{N-1} \end{equation}
where $I^{(j)} $ is a identity operator acting on $j$th block of state and $I_j$ is identity operator acting on $j$th qubit of particular block of state. The operators $P_\pm$ are projection operators given by $(I\pm Z)/2$. The structure of the Hamiltonian shows that the Hamiltonian consists of two local and one local product operators in pauli operator $Z$ making it suitable for implementation in quantum annelears.

 Another possible encoding is one in which each collection $b_t$ encodes a node in a binary system (see Fig.~\ref{fig:encoding}). In this case, for each time we need $K \equiv \lceil \log_2 N \rceil$ qubits. In total we need $\sim N \mathcal{O}(\log_2 N)$ qubits. Since each sequence $b_t$ represents a unique and valid node, $H_\text{valid}(b_t)$ term can be dropped in Eq.~\eqref{eq:cost-function}.
 The state representation look like 
\begin{equation}
	\begin{split}
		|\Psi\rangle &= \bigotimes_{t=0}^{N-1} |\{b_t\}\rangle 
	\end{split}
\end{equation}
where $\{b_t\}$ represents the binary representation of $b_t$. The transformation to Pauli operators given by 
\begin{equation}b_{i,t} \rightarrow  I^{(0)} \otimes \cdots \otimes \mathcal{H}^{(t)}_{\{i\}} \otimes \cdots \otimes I^{(t_\text{max}-1)} \end{equation}
where the operator $ \mathcal{H}^{(t)}_{\{i\}}$ is constructed by replacing $0$ (and $1$) by operators $P_-$ (and $P_+$) in $\{i\}$. The operator $I^{(j)}$ is a identity operator acting on $j$th block of state. In contrast to QUBO encoding, the Hamiltonian is no longer two local operator in pauli operator $Z$, therefore is no longer suitable for the quantum annealers.

Let us examine the number of terms in the Hamiltonian, focusing on the cost function \eqref{eq:cost-function} to estimate its scaling. We begin with the QUBO encoding. According to \eqref{eq:cost-function}, the term  
\[
b_{i,t}b_{j,t+1} = I^{(0)} \otimes \cdots \otimes \mathcal{Q}^{(t)}_i \otimes \mathcal{Q}^{(t+1)}_j \otimes \cdots \otimes I^{(N-1)}
\]  
contains four components: one two-local term, two one-local terms, and the identity operator. Summing over the indices \(i\), \(j\), and \(t\), we find the following contributions:
\begin{itemize}
    \item \((N-1)(N(N-1)/2)\) two-local terms,
    \item \((N-1) \cdot N\) one-local terms,  
    \item and the identity operator.
\end{itemize}
Therefore, the total number of terms in the cost function scales as \(\mathcal{O}(N^3)\).

Next, consider the HOBO encoding. According to \eqref{eq:cost-function}, the term  
\[
\sum_{ij} b_{i,t}b_{j,t+1} = \sum_{ij} I^{(0)} \otimes \cdots \otimes \mathcal{H}^{(t)}_{\{i\}} \otimes \mathcal{H}^{(t+1)}_{\{j\}} \otimes \cdots \otimes I^{(N-1)}
\]  
contains \(\binom{2K}{i}\) \(i\)-local terms for \(i = 0, 1, \dots, 2K\). Summing these contributions over \(i\), \(j\), and \(t\), we obtain  
\[
(N-1)\sum_i \binom{2K}{i} = (N-1)N^2 \sim \mathcal{O}(N^3)
\]  

Thus, the number of terms in the cost function scales in the same order, \(\mathcal{O}(N^3)\), for both encodings. Finally, more rigorous upper bounds as derived in \cite{glos_space-efficient_2022} show that the circuit depths scale as \(\mathcal{O}(N)\) and \(\mathcal{O}(N^2)\), respectively, for these encodings.

In algorithm based on DBG, the $k$-mers are assigned to the edges $e \in \mathcal{E}$, while the $(k-1)$-mers assigned to nodes $i \in \mathcal{V}$, and the goal is to find a Eulerian cycle in the graph. Since $k$-mers can occur multiple times in the original genome, we will allow for the edge to be traced multiple times. This `$k$-mer multiplicity' is encoded in the cost matrix $W_{ij}$ by equating it to large negative value. Therefore, 

\begin{equation}\label{eq:cost-function-01}
H(b) = A_1 \sum_{t=0}^{N-1} H_\text{valid}(b_t) + \sum_{i,j=0}^{N-1}W_{ij}\sum_{t=0}^{N-1} b_{ti}b_{t+1,j}\,
\end{equation}

\FloatBarrier

\subsection{Reference State}

A reference state aids in faster convergence of variational algorithms. In the genome sequencing problem, the circuit is initialized to a reference state that encodes information about the potential source and end nodes. The source node is chosen as a node with no in-degree, while the end node is selected as a node with no out-degree. If no such nodes exist, the source and end nodes are assigned based on the nodes with the minimum in-degree or out-degree weights, respectively. 

\subsection{Ansatz}

\begin{figure*}[ht]
	\centering
	\includegraphics[width=0.8\linewidth]{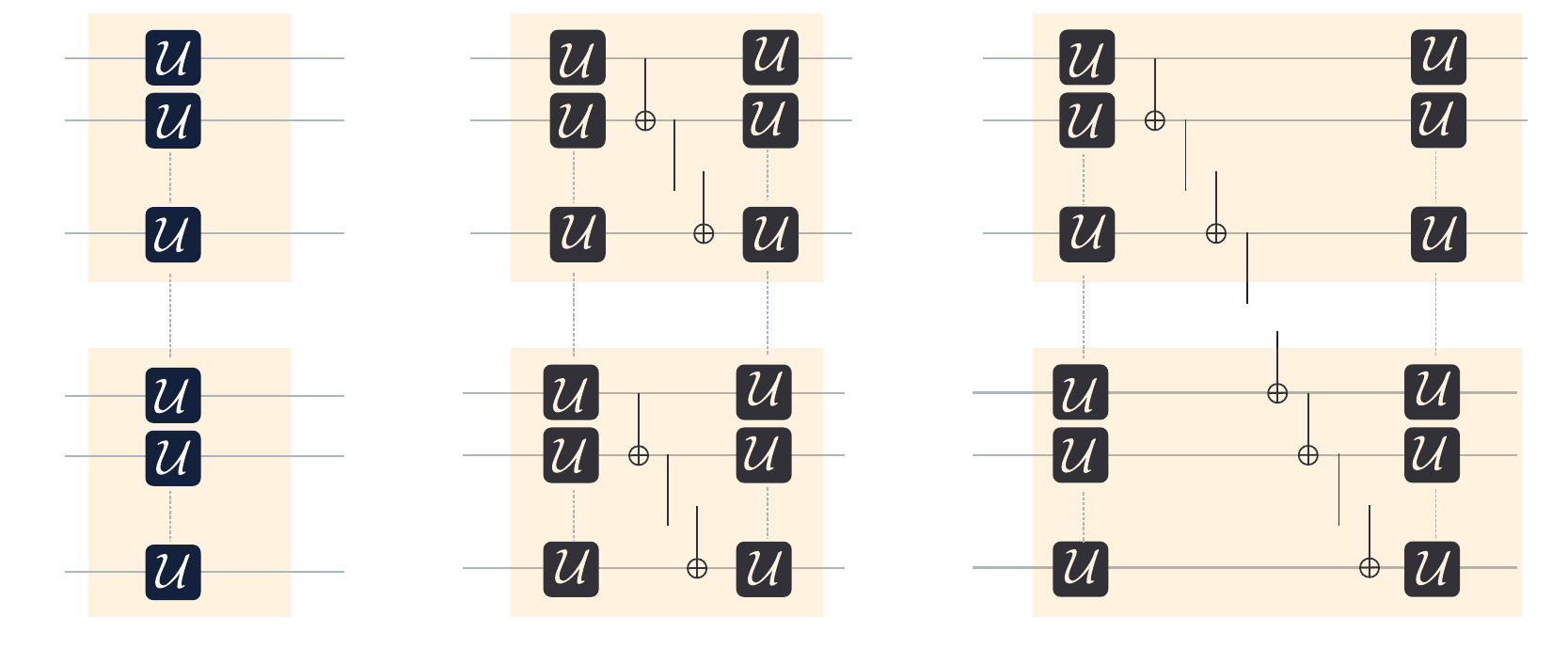}
	\caption{Illustration of three ansatz used in variational quantum design. The quantum gate $\mathcal{U}$ represents single qubit rotation gate with angles $(\theta_x,\theta_y,\theta_z)$. The orange color blocks shows qubits associated with $b_t$ representing particular position or time. \textbf{Left :}The simplest possibility is to consider ansatz which consists of only single qubit rotation gates, therefore, there is no entanglement and the total state $|\Psi\rangle$ remain in product state of individual qubits. \textbf{Centre :} A non-trivial ansatz considers entanglement (CNOT gates) only among qubits ($\log_2 N$ qubits) representing a particular position, in a linear fashion. \textbf{Right :} Ansatz with linear entanglement spanning across all qubits, akin to an EfficentSU2 ansatz.}
	\label{fig:ansatz}
\end{figure*}

To iteratively optimize from a reference state to a target state \( |\psi(\theta_i)\rangle \), we need to define a variational form \( U_V(\theta_i) \). An ansatz capable of exploring the entire Hilbert space would perform poorly due to the exponential dimensionality of the space. Therefore, constructing an efficient ansatz is critical for the success of variational algorithms. In practice, one must balance three key factors: 
\begin{itemize}
    \item reducing the search space for faster convergence $\implies$ speed;
    \item ensuring that the reduced space still contains the actual solution, as excluding it could lead to suboptimal results $\implies$ accuracy;
    \item accounting for deeper circuits as they are susceptible to noise $\implies$ low errors.
    
\end{itemize}
For our purposes, we consider three variational forms with varying degrees of entanglement (see Fig.~\ref{fig:ansatz}). As previously shown\,\cite{PhysRevA.104.062426}, an ansatz with a high degree of entanglement is not always the most efficient choice. The ansatz in the left creates a product state, and the search space is huge. This kind of an ansatz is easy to implement in a hardware as no interaction among qubits is required. The circuit depth is essentially one here. The ansatz shown in the centre of the figure has block entanglements. In this one, there is linear entanglement (ladder shape entanglements) among qubits representing a position in the graph sequence. The ansatz shown in the right has linear entanglement spanning all the qubits. This kind of ansatzes generally tend to have large circuit depths owing to limited qubit connectivity in the hardware. The block ansatz (shown in the centre of the figure) strikes a balance between adequate entanglements and circuit depths.

\FloatBarrier

\section{Results} \label{sec:results}
In this study, we focus on implementing a hybrid classical--quantum approach for \textit{de novo} genome assembly using raw sequencing data from the Sequence Read Archive (SRA) \cite{katz2022sequence} across a diverse set of organisms, including bacteria, fungi, and viruses.

The objective is to assess whether the assembled reads accurately represent the original source organisms. To this end, we analyze single-layout libraries generated using long-read sequencing platforms, including Oxford Nanopore \cite{Wang2021_NanoporeReview} and PacBio \cite{rhoads_pacbio_2015}, and identify the corresponding organisms. The datasets examined in this study include organisms such as Bacillus cereus (SRR12045421), Monkeypox virus (SRR32413059), and African swine fever virus (SRR27477754), among others \cite{katz2022sequence}. A comprehensive list of all datasets is provided in SI Table \ref{tab:sra_datasets_Supp}.

The raw sequencing data from the SRA database underwent pre-processing and quality control to remove low-quality reads and trim sequencing adapters. Quality was assessed based on factors like GC content, per-base sequence quality, and base quality scores (explained in the methodology). For example, the \textit{African Swine Fever virus} (SRR27477754) dataset \cite{katz2022sequence}, with 81.8 Mb and 16,388 reads, passed the quality check and after removing any remaining adapters the read lengths ranged from 400 to 10k bases, as these are long reads. 

\subsection{Graph Generation}
Sequence alignments were performed using the pairwise2 module from the Biopython package in Python \cite{cock_biopython_2009}, which provides efficient algorithms for pairwise sequence comparison, including local alignment methods for assessing similarity between sequencing reads. To speed up the process, we utilized parallel processing with Python's multiprocessing module, which allowed us to handle a large number of sequence pairs simultaneously, leveraging multiple CPU cores for faster computation. Next, we constructed OLC graphs, where each read represented a node. The edges between nodes were determined based on the alignment scores between pairs of sequences, achieved by aligning the suffix of one read with the prefix of the next. To find the best overlaps, we used local alignments. We also removed exact duplicates (those with $\geq 99.9\%$ identity) and ignored overlaps with more than 4 mismatches, as our goal was to prioritize cleaner, more reliable overlaps rather than filling gaps with low-quality alignments. This process allowed us to build a graph by connecting reads based on their shared overlapping regions. We applied a threshold to filter out low-quality alignments, ensuring that only reads with strong overlaps were considered, which helped keep the graph smaller and more manageable. This filtering was important given the current limitations of qubits in the present-day IBM quantum hardware, which has a maximum capacity of 156 qubits. The filtering can be relaxed to accommodate larger graphs as QPUs with more qubits become available.

\begin{figure*}
	\centering
	\includegraphics[width=1.0\linewidth]{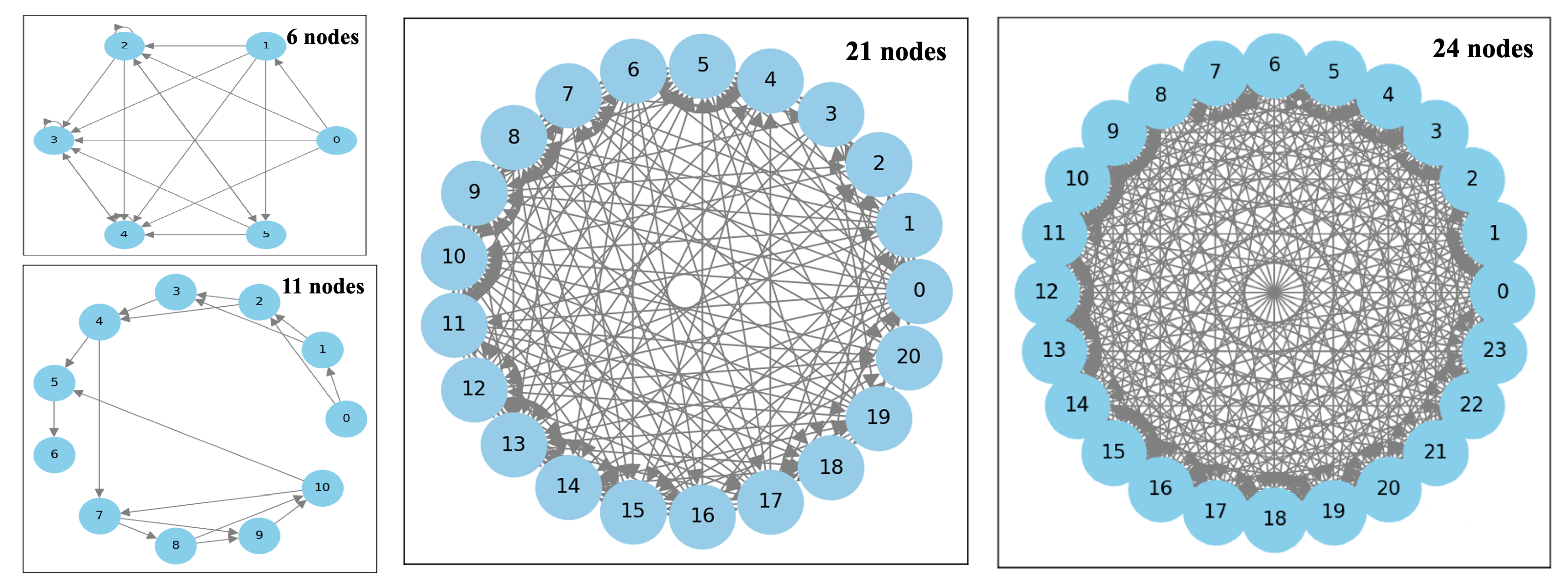}
	\captionsetup{justification=justified,singlelinecheck=false}
	\caption{Directed graphs representing partial genome assembly graphs containing 6, 11, 21, and 24 nodes, respectively, generated from sequencing reads of different organisms after filtering and removal of low-quality alignments}
	\label{fig:directed_graphs}
\end{figure*}
\FloatBarrier

In Fig. \ref{fig:directed_graphs}, graphs generated from the\textit{ African Swine Fever virus} (SRR27477754) reads are represented with 21 and 24 nodes. After graph construction, we identified the Hamiltonian path, which was then used to assemble a contig of the genome sequence. To validate the accuracy of our graph-based approach, we compared it against classical Hamiltonian path solutions using dynamic programming. The resulting path was processed to form the contig, which was then queried against a database using BLASTn to identify the organism \cite{altschul1990basic}.
The organism identification results from classical methods, presented in SI Table \ref{tab:supp_classical_organism_identification}, show that for smaller graphs (4 - 6 nodes), while the original organism is identified (SI Table \ref{tab:sra_datasets_Supp}), other strains or genera also appear due to the smaller size of the generated contigs. However, as the graph size increases (7+ nodes), the organism's identification becomes more accurate. In particular, larger graphs (beyond 7 nodes) yielded precise identifications, with the exception of the 9-node case, where a different strain (USA\_CA0103) of the monkey virus (\ref{tab:supp_classical_organism_identification}) was identified rather than the Kenyan strain (\ref{tab:sra_datasets_Supp}). Overall, even with smaller graph representations, we were able to accurately identify the organism and determine its taxonomic status in a \textit{de novo} context, achieving improved query coverage and \% identity. This method can also aid in identifying previously unknown organisms by generating subsets of the graph and exploring Hamiltonian path based solutions within them.

\subsection{Analyzing results from the simple CVaR-VQE} \label{sec:result-analysis}
\begin{figure*}
	\centering
	\includegraphics[width=0.8\linewidth]{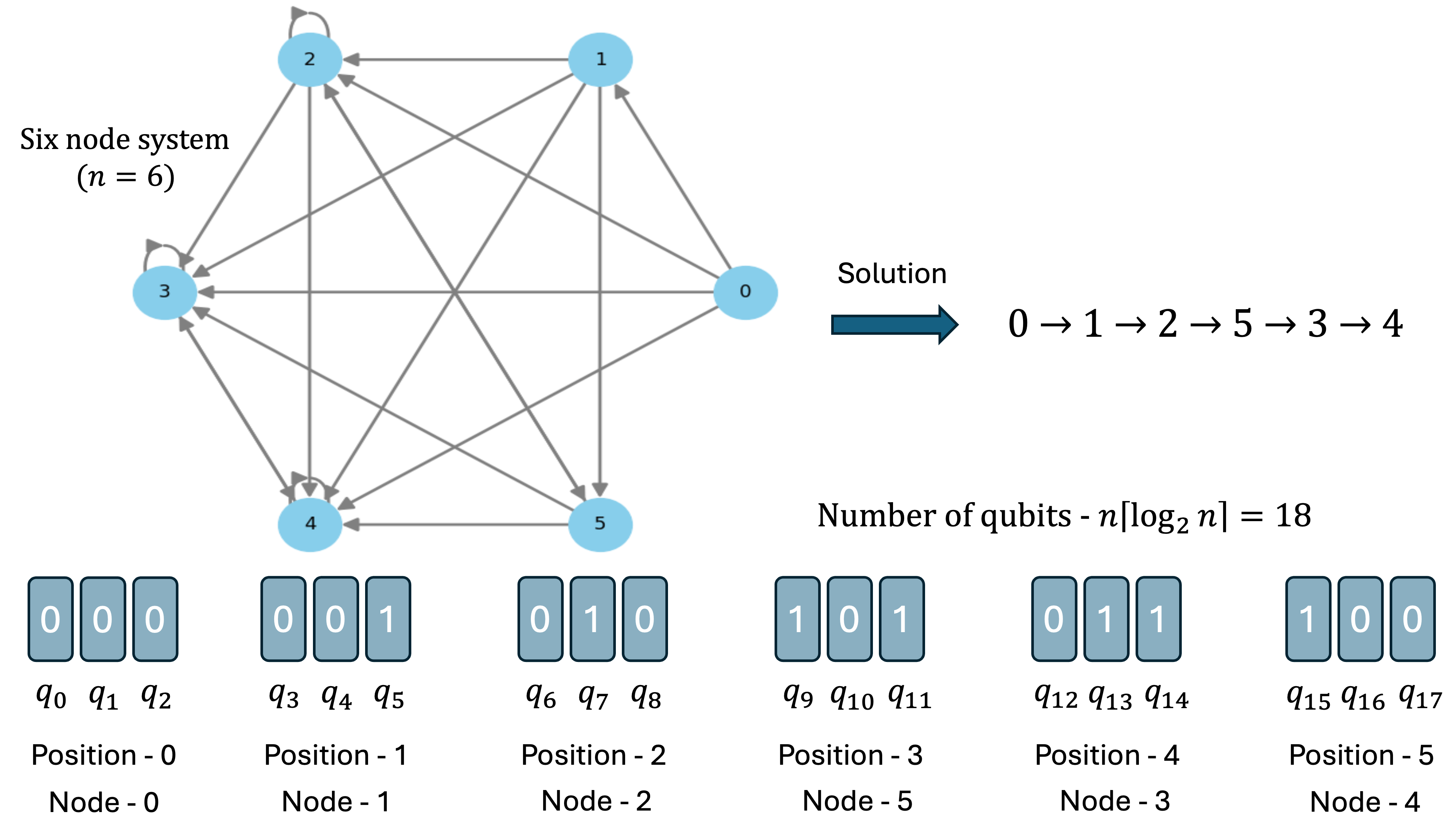}
	\caption{Representative solution for a 6-node system}
	\label{fig:bitsting_decode}
\end{figure*}

Fig.\ref{fig:bitsting_decode} gives us a representative solution for the 6-node system. Each position is represented by $3$ qubits and the output bitstrings assigns nodes to these positions. Fig.\ref{fig:bitsting_decode} has the illustration.

\FloatBarrier

As we scaled up the problem to higher number of graph nodes with denser connectivity, such as in the range 18-24 nodes (uses 90-120 qubits), we noticed two main kinds of issues with the bitstring outputs. 

Error type I: Anomalies (repetitions and invalid nodes) - 
In the resulting sequence, it is sometimes observed that incorrect nodes appear as part of the solution. For instance, a sequence may contain a repeated node, or an invalid node that does not exist in the original graph. As an example, for a graph with 12 nodes, the decoded bitstring output may correspond to a path in which node 2 appears twice (indicating a repetition), or a node comes out as 14, which lies outside the range of valid nodes. It should be noted that a well-constructed cost function should inherently account for repetitions through its constraints.

In the case of invalid nodes, their occurrence is likewise a consequence of the chosen construction, for example whether one uses QUBO or HOBO encodings. In such formulations, nodes are represented using bitstrings, whose length is not always an exact power of two ($2^n$). For instance, for a $12$-node graph, with HOBO encoding, four qubits are required to encode each node, allowing representations from $0000$ to $1111$. If the valid nodes correspond to bitstrings $0000$ - $1011$ (representing nodes $0$ to $11$), then the bitstrings $1100$, $1101$, $1110$ and $1111$ correspond to invalid nodes. During post-processing of the results, when groups of four bits are interpreted as node labels, such invalid bitstrings may appear due to noise or optimization output errors. This can lead to anomalies in the extracted sequence, which must be handled appropriately. In particular, these scenarios are very significant when running on real hardware, where readout errors or other kind of errors that disturb even singular qubits can manifest as a repetition or invalid node. 

It is important to note the sources of these errors: anomalies can arise from imperfect cost function design and noise-related sources such as readout errors or decoherence.

\begin{figure*}[ht]
	\centering
	\includegraphics[width=0.95\linewidth]{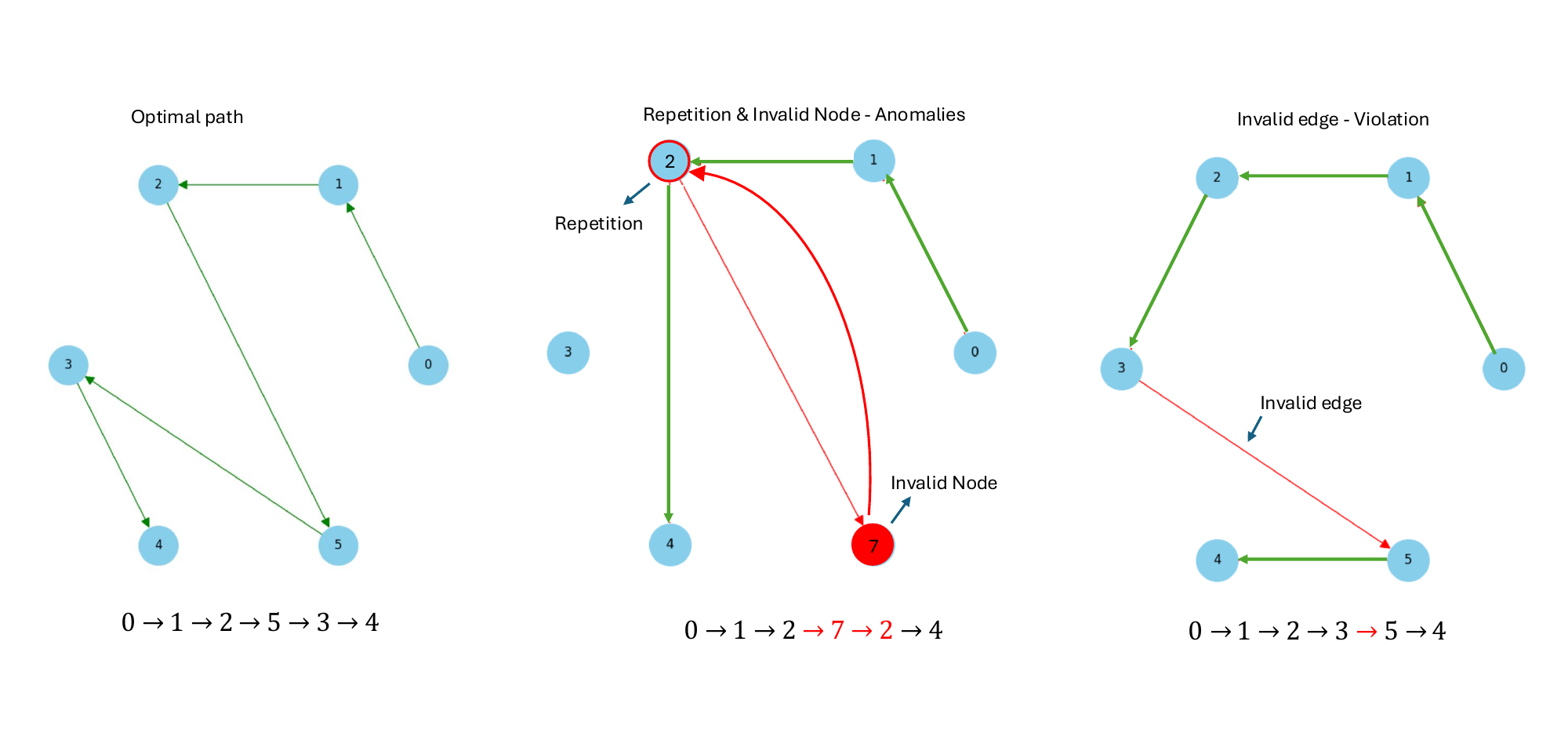}
	\caption{Types of error in bitstrings and corresponding configuration issues}
	\label{fig:errors}
\end{figure*}

Error type II: Violations (invalid edges in the path) 
As mentioned in Supplementary Section~\ref{method:VQA}, we interpret the path from the sequence. In the example sequence $\left[0,1,2,3,5,4\right]$, the path is 
$$0\rightarrow 1\rightarrow 2\rightarrow 3\rightarrow 5\rightarrow 4\,$$ Comparing this path with the edges on the original graph, we see that the edge $3\rightarrow 5$ that is not originally present in the graph, appears. These are violations and represent missteps in the overall final path. Fig. \ref{fig:errors} gives an illustration of the types of errors.

\FloatBarrier
Excluding sources such as noise, and problem-specific sources such as imperfect cost function design and suboptimal optimization path traversal, violations are a natural consequence of the iterative process of optimization with VQE. This occurs in part because information about existing graph edges is not explicitly hardwired as hard constraints. Instead, the objective function (refer the Hamiltonian in (\ref{eq:cost-function-01})) incentivizes transitions corresponding to valid edges and penalizes transitions corresponding to non-existent edges through the adjacency matrix. 
During optimization, intermediate solution states (sequences) are expected to iteratively evolve from states with a higher number of violations towards states with a fewer number of violations, ideally zero. While post-processing schemes to handle residual violations are necessary, any post-processing approach that guides the optimization toward more favourable traversal paths can also improve the final solution by reducing the number of violations.
The bitstring recovery mechanism described in Section \ref{sec:bit_rec} is designed to achieve this effect. This becomes particularly important for problems with a larger number of nodes, where we observed sub-optimal convergence behavior.


\subsection{VQE runs in simulator and hardware}\label{vqeruns}
We tested our formulation on several sequences, ranging from $4$ nodes up to $24$ nodes. As mentioned above, the number of qubits required is $\sim N \mathcal{O}(\log_2 N)$. With $24$ nodes this equals $120$ qubits, which is close to the number of qubits available ($156$) on the latest IBM machines. Fig.~\ref{fig:directed_graphs}) gives the directed graphs of some of the sequences. While trying to find out the optimal path for each of the sequences, we tried out several combinations of ansatz, optimizers, starting points, and initial parameters on a CVaR-VQE set up \cite{Barkoutsos-cVar}. Based on our observations, we realized that in a simulator setting, the SPSA optimizer works the best. This is contrary to the observations made in \cite{opti_analysis}, where COBYLA seems to have outperformed SPSA. This difference may be due to the specific nature of our problem formulation and the associated optimization landscape, which could favour the behaviour of SPSA. We used parameters provided by semi-converged VQE runs in the simulators as initial points for the hardware runs. In the hardware, we used SPSA as well as COBYLA optimizers. 

In most cases, the block ansatz shown in Fig. \ref{fig:ansatz} (centre) yielded the best performance. This ansatz provides an effective balance between entangling operations and single-qubit rotations, while keeping the overall circuit depth relatively low. To further reduce the optimization burden on the VQE, we also explored fixing the starting and ending graph positions. This was implemented by applying appropriate $X$ gates to the first and last $\log_2 N$ qubits. The penalty factor $A_1$ was tuned empirically using a trial-and-error approach.

For highly connected graphs, such as the $21$ and $24$ node systems, we additionally applied differential weighting to non-existent adjacencies. This penalization discourages the VQE from producing sequences corresponding to non-overlapping reads. Beyond these fine-tuning steps, we actively incorporated the bitstring recovery technique described in Section \ref{sec:bit_rec} to further improve solution quality and optimizer stability.

\begin{figure}[!htbp]
\centering
\captionsetup[subfloat]{labelformat=parens}

\subfloat[]{%
  \includegraphics[width=0.45\linewidth]{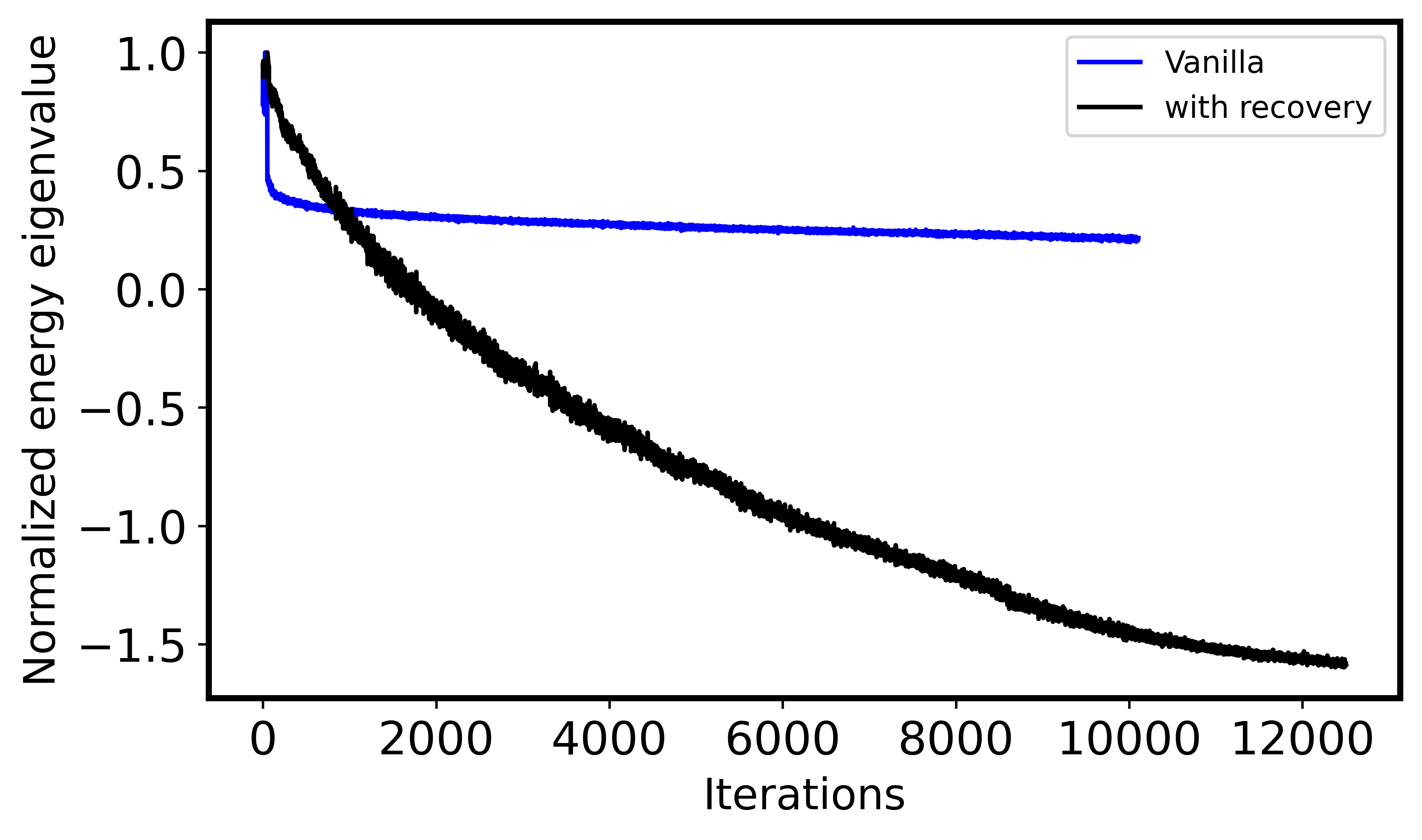}%
  \label{fig:normalized_prop_a}%
}\hfill
\subfloat[]{%
  \includegraphics[width=0.53\linewidth]{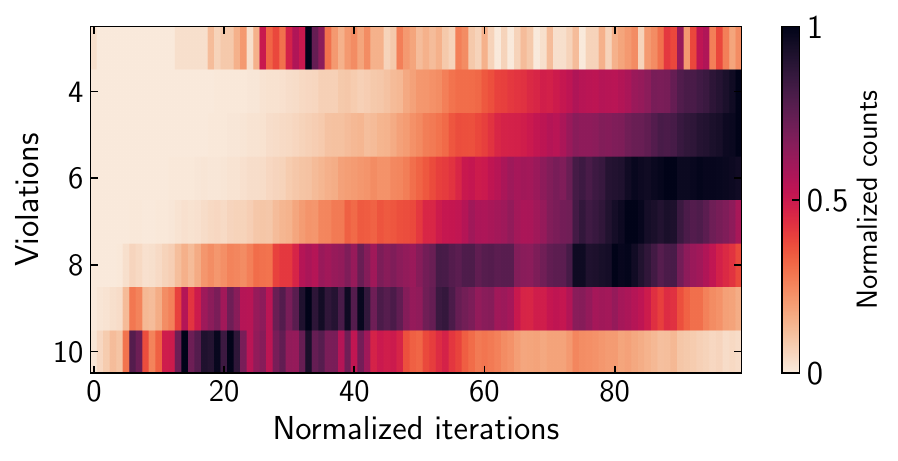}%
  \label{fig:normalized_prop_b}%
}

\caption{Simulation result obtained using an MPS simulator for a 21-node graph. (a) Normalized energy of the VQE Hamiltonian as a function of optimization iterations, showing convergence toward the ground-state energy, obtained using 'Vanilla' and the recovery technique. (b) Evolution of constraint violation during the optimization (using VQE with recovery): as the iteration increases and the system approaches the ground state, the number of violations decreases. The color scale indicates the number of sampled solutions exhibiting a given number of violations at each iterations. The number of iteration is normalized to 100 and counts to 1 for better visibility.}
\label{fig:normalized_prop}
\end{figure}

\begin{table*}
\centering
\caption{\label{table:simulator_results} Results from simulator with starting and ending nodes provided. (NR - Not Required)}
\begin{tabular}{*5c}
    \toprule
    \multicolumn{3}{c}{} & \multicolumn{2}{c}{Violations} \\
    \cmidrule(rl){4-5}
    No. of nodes & Qubit used & Hamiltonian terms & Vanilla  & With recovery \\
    \midrule
    \rowcolor{gray!10}
    4  & 8   & 48    & 0 & NR \\ 
    6  & 18  & 538   & 0 & NR \\ 
    \rowcolor{gray!10}
    7  & 21  & 1079  & 0 & NR \\ 
    9  & 36  & 8236  & 0 & NR \\ 
        \rowcolor{gray!10}
    18 & 90  & 82312 & 2 & 0  \\ 
        \rowcolor{gray!10}
    21 & 105 & 202462 & 5 &  2 \\ 
    24 & 120 & 53415 & 5 & 2 \\ 
    \bottomrule
\end{tabular}
\end{table*}
Table\,\ref{table:simulator_results} summarizes the results from MPS simulator\,\cite{aer_simulator} for graphs where the start and end nodes were provided. The simulator was run in two different setups. In the first setup, we did not employ any recovery mechanism, and call it the `Vanilla' setup. In the second setup we employed the bitstring recovery mechanism in every iteration of the VQE. The `Vanilla' setup was adequate until the $11$ node system and no other recovery techniques were required (mentioned as NR in Table\,\ref{table:simulator_results}). For the $18$ node system, we were able to obtain the optimal sequence with the bitstring recovery mechanism employed. However, for the $24$ node system, we could get it down to $2$ violations from the $5$ violations we had with the `Vanilla' setup. Fig.\,\ref{fig:normalized_prop}a shows the VQE output convergence plot for the $21$ node system using both the 'Vanilla' setup and with the recovery technique implemented at every iteration. It can be seen that VQE with the recovery technique had converged better to a global minima, while the 'Vanilla' setup stayed at a local minima till 10000 itrations. The SPSA optimizer was used in this case with $4000$ shots per iteration.


As the iterations progress, the proportion of solutions exhibiting higher number of violations decreases, while those with fewer violations increase. This trend is illustrated in Fig.\,\ref{fig:normalized_prop}b, which shows the normalized counts of solutions with a given number of violations plotted against the normalized iteration count (scaled to $100$). The normalization of solution counts was performed separately for each violation level, using the maximum number of solutions observed for that violation level across all iterations as the normalization factor. For instance (refer Fig. \ref{fig:normalized_prop}b), solutions with $10$ violations reached their maximum count around iteration $20$, whereas those with $9$ violations peaked around iteration $30$.

It should be noted that this normalization does not imply that the absolute number of highly violating solutions is smaller than that of less violating ones; rather, it demonstrates that, relative to their respective maxima, the proportion of solutions with fewer violations increases over successive iterations.

\FloatBarrier

Table \ref{table:hardware_results} summarizes the results for graphs where the start and end nodes were not specified. The experiments were conducted as described earlier, with the hardware runs executed for $50$ iterations and warm-started using the optimal parameters obtained from the simulator. To mitigate sampling errors in the hardware runs, Mthree error mitigation \cite{mthree} was applied to the bitstrings obtained at each iteration. Further, the bitstring recovery mechanism mentioned above was applied only for graphs with $18$ nodes and above.  For smaller graphs, a single violation was typically observed, corresponding to a cyclic ordering error, wherein the node sequence is rotated while preserving the correct set of edges. Notably, for the $7$- and $11$-node graphs, the hardware was able to rectify this cyclic error. For larger and more densely connected graphs with $21$ and $24$ nodes, $3$ and $4$ violations were observed in the hardware, respectively. This shows that the bitstring recovery mechanism actively works to promote better traversal of the optimizer in the solution space, leading to a decrease in the observed violations even with only $50$ iterations on hardware for the $21$ and $24$ node system.


\begin{table*}
\centering
\caption{\label{table:hardware_results} Results from simulator ($4000-10000$ iterations) and IBMQ hardware (warm-started, 50 iterations) with bitstring recovery mechanism but without starting and ending nodes specified}
\rowcolors{-1}{}{gray!10}
\begin{tabular}{*4c}
\toprule
 \multicolumn{2}{c}{} & \multicolumn{2}{c}{Violations} \\
    \cmidrule(rl){3-4}
No. of nodes & Two qubit depth of transpiled circuit & Simulator  & IBM Quantum \\
    \midrule
    4  & 4   & 0 & 0 \\ 
    6  & 12  & 0 & 0 \\ 
    7  & 14  & 1 & 0 \\ 
    9  & 27  & 1 & 1 \\ 
    18 & 72  & 0 & 0  \\ 
    21 & 84  & 2 & 3 \\ 
    24 & 130 & 3 & 4 \\ 
    \bottomrule
\end{tabular}
\end{table*}
\subsection{Contig generation}

Once a valid path is obtained from the OLC graph using either quantum or classical solution methods, the ordered sequence of reads is used to construct a contig, representing a contiguous assembled segment of the genome. Given current limitations of quantum hardware, OLC graphs in this study were constructed from reduced subsets of full read datasets. The goal of this work is not full genome reconstruction, but rather the generation of informative contigs that enable organism or taxonomic group identification in a \textit{de novo} assembly setting where the source organism is unknown.

To validate the proposed approach, OLC graphs were generated from sequencing reads derived from organisms with known reference genomes, as listed in SI Table~\ref{tab:sra_datasets_Supp}. These datasets served as benchmarks to assess whether the pipeline could accurately recover organism identity. For each dataset, OLC graphs were constructed and solved, and the resulting paths were used to assemble contigs. The assembled contigs were queried against the NCBI nucleotide database using BLASTn~\cite{altschul1990basic}, with a minimum word size of 16 and searches restricted to complete organism genome sequences.

Organism identity was inferred based on BLASTn query coverage, indicating the fraction of the contig aligned to reference sequences, and percentage sequence identity, reflecting nucleotide-level similarity within the aligned regions. The highest-scoring match was considered the most likely source organism. A summary of these results is presented in Table~\ref{tab:organism_identification}, which reports the read datasets, corresponding OLC graph sizes, and organism identification outcomes.

Table~\ref{tab:organism_identification} primarily highlights results obtained from quantum hardware solutions applied to the OLC graphs using the proposed quantum workflow. Expected Hamiltonian paths were obtained classically using a dynamic programming approach, as described in Supplementary Section~\ref{method:dynamic}, and served as the reference for comparison. If the quantum-derived path matched the expected Hamiltonian path, the violation count was recorded as zero, otherwise, missing or invalid edges were counted as violations.

Graphs of size 4, 6, 7 and 18 exhibit zero violations, indicating correct recovery of the Hamiltonian path. In these cases, organism predictions derived from the assembled contigs are accurate and highlighted in red (Table \ref{tab:organism_identification}). In smaller graph systems, where contigs are constructed from fewer reads, the assembled sequences may align to multiple closely related organisms. For example, in the 4-node and 6-node systems, the original source organisms are \textit{Bacillus cereus} and the \textit{hDenV2/Bangladesh/2023} strain, respectively (Table \ref{tab:sra_datasets_Supp}). Although correctly identified with high query coverage and sequence identity, additional organisms with comparable scores are also observed. This reduced specificity can be due to the limited number of reads contributing to contig generation.

In contrast, larger graph systems (18, 21 and 24 nodes) yield more specific organism identification, with the correct source organism consistently exhibiting higher query coverage and percentage identity than alternative matches (Table \ref{tab:organism_identification}). For the 9, 21 and 24-node systems, quantum hardware solutions exhibit violation counts of 1, 3 and 4, respectively. Despite the presence of these violations, the assembled contigs still enable correct organism identification, albeit with reduced query coverage compared to classical solutions. For instance, in the 24-node system, the classical solution achieves 78\% query coverage with 94.26\% identity for the \textit{African swine fever virus isolate NAM/P1/1995}, while the quantum solution yields the same organism identification with a reduced query coverage of 75\%. A similar trend is observed for the 21-node system, where three violations correspond to a decrease in query coverage from 56\% to 49\%, while preserving correct organism identification (Table \ref{tab:organism_identification}).

Overall, these results demonstrate that quantum-assisted OLC graph solving can produce biologically meaningful contigs and accurate organism identification, even when applied to limited read subsets and in the presence of constraint violations, highlighting the potential of quantum-enhanced workflows for \textit{de novo} genome assembly.

{\sloppy
\begin{table*}
\centering
\caption{Summary of read datasets, graph size, violation counts, and organism identification from assembled contigs obtained using quantum hardware.}

\label{tab:organism_identification}
\renewcommand{\arraystretch}{1.25}

\begin{tabularx}{\textwidth}{l c c X c c}
\toprule

\multirow{2}{*}{\textbf{Read dataset}} &
\multirow{2}{*}{\textbf{Graph nodes}} &
\multirow{2}{*}{\textbf{Violations}} &
\multicolumn{3}{c}{\textbf{Organism identified from the assembled contig}} \\
\cmidrule(lr){4-6}
 & & & \textbf{Name} & \textbf{Query coverage (\%)} & \textbf{Identity (\%)} \\
\midrule

\rowcolor{blue!10}
SRR12045421 & 4 & 0 &
\textit{Bacillus thuringiensis} & 100 & 96.19 \\
\rowcolor{blue!10}
 & & &
\textcolor{red}{\textit{Bacillus cereus}} & 99 & 96.07 \\
\midrule

\rowcolor{blue!10}
SRR32137766 & 6 & 0 &
\textit{Dengue virus type 2 isolate Phuket 2020} & 80 & 89.07 \\
\rowcolor{blue!10}
 & & &
\textcolor{red}{\textit{hDenV2/Bangladesh/NRICh-DS028/2023}} & 79 & 88.62 \\
\midrule

\rowcolor{blue!10}
SRR13062400 & 7 & 0 &
\textcolor{red}{\textit{Bacteroides intestinalis}} & 95 & 93.66 \\
\midrule

\rowcolor{blue!10}
\multirow{2}{*}{SRR32413059} &
\multirow{2}{*}{9} &
1 (Quantum) &
\textit{Monkeypox virus isolate USA\_CA0103} & 100 & 95.87 \\
\rowcolor{gray!15}
 & & 0 (Classical) &
\textit{Monkeypox virus isolate USA\_CA0103} & 100 & 96.5 \\
\midrule

\multirow{2}{*}{\cellcolor{blue!10}SRR32006043} &
\multirow{2}{*}{\cellcolor{blue!10}18} &
\multirow{2}{*}{\cellcolor{blue!10}0} &
\cellcolor{blue!10}\textit{Influenza A (Antarctica\_Tern/H5N1)} &
\cellcolor{blue!10}83 &
\cellcolor{blue!10}91.69 \\
\cellcolor{blue!10} & \cellcolor{blue!10} & \cellcolor{blue!10} &
\cellcolor{blue!10}\textcolor{red}{\textit{Influenza A (Brown\_Skua/H5N1)}} &
\cellcolor{blue!10}83 &
\cellcolor{blue!10}91.69 \\
\midrule

\rowcolor{blue!10}
\multirow{2}{*}{SRR27477754} &
\multirow{2}{*}{21} &
3 (Quantum) &
\textcolor{red}{\textit{African swine fever virus isolate NAM/P1/1995}} & 49 & 91.81 \\
\rowcolor{gray!15}
 & & 0 (Classical) &
\textcolor{red}{\textit{African swine fever virus isolate NAM/P1/1995}} & 56 & 91.81 \\
\midrule

\rowcolor{blue!10}
\multirow{2}{*}{SRR27477754} &
\multirow{2}{*}{24} &
4 (Quantum) &
\textcolor{red}{\textit{African swine fever virus isolate NAM/P1/1995}} & 75 & 94.26 \\
\rowcolor{gray!15}
 & & 0 (Classical) &
\textcolor{red}{\textit{African swine fever virus isolate NAM/P1/1995}} & 78 & 94.26 \\
\bottomrule
\end{tabularx}
\vspace{0.5em}
\begin{minipage}{\textwidth}
\footnotesize
\textbf{*} Blue rows denote quantum hardware solutions along with their associated violation counts, whereas grey rows denote the corresponding classical solutions with zero violations. Correctly predicted organisms are highlighted in red.
\end{minipage}
\end{table*}
}

\section*{Conclusion}
In this paper, we present a quantum-assisted framework for \textit{de novo} genome assembly by formulating the Hamiltonian path problem arising in OLC and DBG graphs. Classically, solving the Hamiltonian path problem exhibits factorial complexity, $\mathcal{O}(N \times N!)$, making it intractable for large assembly graphs. While prior quantum formulations reduced the encoding cost to $\mathcal{O}(N^2)$ qubits, we introduce the HOBO formulation, which further reduces the qubit requirement to $\mathcal{O}(N \log_2 N)$. We employ a conditional value-at-risk (CVaR) based variational quantum eigensolver (VQE), together with a novel bitstring recovery technique, to confine the optimizer's exploration of the solution space to regions corresponding to valid assembly paths. Experiments performed on both simulators and IBM quantum hardware successfully recovered optimal paths for graphs containing up to 18 nodes, while for highly connected graphs with 21 and 24 nodes only limited constraint violations were observed.

Despite these violations, downstream contig construction and organism identification analyses showed that the quantum-derived paths remained biologically meaningful. In particular, accurate organism identification was achieved using subsets of assembly graphs, with reductions in query coverage observed only when constraint violations occurred. These results suggest that Hamiltonian path problems on subsets of assembly graphs can be addressed using quantum formulations with logarithmic encoding overhead, supporting reliable organism identification within \textit{de novo} assembly workflows. While the present study is intentionally limited to organism identification using subsets of sequencing reads due to current quantum hardware constraints, the proposed formulation provides a scalable framework for encoding and solving increasingly complex assembly graphs. As quantum hardware continues to advance, this approach may extend beyond partial assemblies toward full-scale \textit{de novo} genome reconstruction at the whole-genome level.

\acknowledgments
This work was supported, in part, by a grant from Mphasis to the Centre for Quantum Information, Communication, and Computing (CQuICC) at IIT Madras.

\FloatBarrier

\putbib
\end{bibunit}


\clearpage

\setcounter{equation}{0}
\setcounter{figure}{0}
\setcounter{table}{0}
\setcounter{section}{0}

\renewcommand{\theequation}{S\arabic{equation}}
\renewcommand{\thefigure}{S\arabic{figure}}
\renewcommand{\thetable}{S\arabic{table}}
\renewcommand{\thesection}{S\arabic{section}}
\renewcommand{\theHequation}{S\arabic{equation}}
\renewcommand{\theHfigure}{S\arabic{figure}}
\renewcommand{\theHtable}{S\arabic{table}}
\renewcommand{\theHsection}{S\arabic{section}}

\begin{bibunit}

\section{Supplementary Methodology}\label{supp:method}
We present a detailed description of our computational pipeline below. The workflow begins with classical pre-processing, followed by graph construction and Hamiltonian path identification using classical methods for benchmarking purposes. We then describe the implementation of variational quantum algorithms employed to address the Hamiltonian path problem within a quantum computing framework. The pipeline concludes with contig generation and organism identification. A schematic overview of the complete workflow is provided in Scheme \ref{sch:workflow}.

\subsection{Classical Pre-processing}
Raw FASTQ files were obtained from Sequence Read Archive (SRA) datasets \cite{katz2022sequence} for multiple organisms with genome sizes ranging from approximately 5 to 80 Mb. The reads were subsequently preprocessed through quality control assessment and adapter trimming to ensure high-quality input for genome assembly.

\subsubsection{\textbf{Quality Control}}
To begin with the quality control (QC) of the raw FASTQ reads, various parameters of the reads are examined to identify and remove low-quality sequences. This ensures that only high-fidelity reads are retained for subsequent genome assembly and analysis.

\textbf{Quality Scores:}
Each base in a sequencing read is assigned a quality score that reflects the probability of an incorrect base call during sequencing. These scores are computed using the Phred scale, where the quality score ($Q$) is defined as:

\begin{equation}
Q = -10 \cdot \log_{10}(P)
\label{eq:phred}
\end{equation}

Here, $P$ represents the probability of an error in the sequencing process for a particular base. The scores are encoded in ASCII format in the FASTQ file, typically ranging from 0 (lowest) to 42 (highest).  0-20 as poor quality reads, 21-28 as medium-quality and 29-42 as best quality reads. Only the reads belonging to the best-quality category were retained for downstream analysis.\\

\textbf{GC Content:}
GC content represents a fundamental property of DNA sequences that affects genome stability, structure, and evolution. Regions with higher GC content exhibit stronger base pairing due to the three hydrogen bonds between guanine and cytosine compared to the two between adenine and thymine.  

The GC percentage across all reads was calculated using Equation~\ref{eq:gc_content}:

\begin{equation}
\mathrm{GC}\% = \frac{\text{Count(G + C)}}{\text{Count(A + T + G + C)}} \times 100
\label{eq:gc_content}
\end{equation}

GC content also influences sequencing efficiency and accuracy, and abnormal GC distributions may indicate contamination or biased amplification.\\

\textbf{Base Sequence Quality:}
The quality scores of each base position across all reads are plotted as \textit{Quality Score (Y-axis)} vs. \textit{Base Position in Read (X-axis)}. This visualization helps identify regions of low sequencing accuracy. Reads with a quality score below 28 at any position were discarded from further analysis.\\

\textbf{Per-Sequence Quality Scores:} 
This metric plots the \textit{Mean Sequence Quality Score (Y-axis)} against the \textit{Sequence Index (X-axis)}. It enables detection of reads with overall poor quality, allowing targeted filtering of such sequences.\\

\textbf{Per-Base Sequence Content:} 
The percentage of each nucleotide (\textit{A, T, C, G}) is plotted as a function of the base position within each read. This provides insights into compositional biases or irregularities within the sequencing data.\\

\textbf{Per-Sequence GC Content:}
The distribution of GC content across reads is used to assess whether the sample's GC profile is consistent with expectations for the organism being sequenced. Deviations from the expected GC distribution may indicate sequencing artifacts or contamination.

\subsubsection{\textbf{Adapter Trimming}}
Following quality control, adapter sequences introduced during library preparation were removed. These adapters facilitate binding of reads to the sequencing flow cell but must be trimmed prior to assembly to avoid artificial overlaps and chimeric contigs. Several commonly used adapter sequences, listed in Supplementary Information (SI) Table~\ref{table:adapters}, were obtained from the FASTQC adapter reference list \cite{fastqc_adapter_list}.

\subsection{Graph Construction}
The next step involves constructing genome assembly graphs from the processed reads and identifying optimal paths within these graphs, which determine the order in which reads are connected to form the assembled genome. Short-read data are generally represented using De Bruijn graphs (DBG), whereas long-read data are modeled using Overlap-Layout-Consensus (OLC) graphs.

For long-read data, the OLC graph was constructed by aligning the suffix of one read with the prefix of the next, with overlap scores \( \mathcal{S}(r_i, r_{i+1}) \) computed as:
\begin{equation}
\mathcal{S}(r_i, r_{i+1}) = \sum_{k} \sigma(a_k, b_k),
\end{equation}
where \( a_k \) and \( b_k \) are the aligned positions of reads \( r_i \) and \( r_{i+1} \), and \( \sigma() \) is the alignment scoring function. The assembly problem was formulated as Hamiltonian path problem, aiming to maximize the total overlap score. Quantum computing algorithms, namely, variational solvers were used to solve the Hamiltonian path problem in quantum simulators as well as IBM quantum hardware. The identified optimal paths (from the quantum runs) were then used to generate contigs and the final genome sequences. For benchmarking, classical dynamic programming was applied to solve the same OLC graphs, enabling a comparison between quantum and classical methods in genome assembly.

For short-read data, genome assembly was formulated as an Eulerian path problem using a DBG. A DBG was constructed by first generating all possible \( k \)-mers from the reads, where each \( k \)-mer is represented by a directed edge between two nodes corresponding to its \( (k-1) \)-mers. This method efficiently captures overlaps between adjacent \( (k-1) \)-mers, with the choice of \( k \)-mer length affecting assembly accuracy and contiguity. The DBG, with \( n \) nodes and \( e \) edges, was treated as a graph traversal problem. Quantum algorithms, using a Hamiltonian objective function, identified optimal paths through the graph, with state vectors representing the most probable genome reconstructions. In parallel, classical methods were also employed for comparison. 

\subsection{Hamiltonian Path Identification Using Dynamic Programming - For Classical Comparison}\label{method:dynamic}

Hamiltonian path identification was performed using a dynamic programming approach with memoization over a directed graph $G = (V, E)$, where $|V| = N$ \cite{gfg_hamiltonian_dp}. Each node represents a vertex, and each directed edge $(u, v) \in E$ defines a possible transition. The algorithm explores all possible node sequences such that each vertex is visited at most once, using a recursive formulation with state caching to reduce redundancy. The dynamic programming approach defines a state function $dp(v, S)$, which stores all possible paths that end at node $v$ and have already visited the set of nodes represented by the bitmask $S$. Here, $S$ is a binary integer in which each bit position indicates whether a node has been visited ($1$) or not ($0$). The recursive relation can be expressed as 
\begin{equation}
dp(v, S) = \bigcup_{u \in \text{Pred}(v)} \{ dp(u, S \setminus \{v\}) + v \},
\end{equation}

where $\text{Pred}(v)$ represents all the predecessor nodes connected to $v$. The base case is $dp(v, \{v\}) = [v]$, meaning that a path containing only a single node $v$ is trivially valid. Bitmasking enables constant-time checks for whether a node is already included in a given path using the logical operation $(S \, \& \, (1 \ll i)) \neq 0$, where $i$ is the node index.
To optimize performance, a memoization strategy using the \texttt{lru\_cache} function stores intermediate results for each $(v, S)$ state, significantly reducing redundant subproblem computations.  Although the theoretical time complexity of the Hamiltonian path problem is exponential, scaling as $O(N^2 2^N)$ due to the enumeration of all vertex subsets, practical execution is substantially accelerated through memoization and graph sparsity.

\subsection{Workflow of Variational Quantum Algorithms} \label{method:VQA}

Every variational algorithm has three major components - the ansatz, the cost function, and the optimizer. The construction and interplay of these components are key factors that influence the success of the algorithm \cite{Peruzzo}. The choices of these components and their construction are discussed in detail in Section \ref{sec:HOBO}. 

The workflow of any variational algorithm is as follows: an ansatz is a `general state' created using a parametrized quantum circuit that represents the set of possible solutions that can represent the solution of the problem at hand. The parameters from the quantum circuit work together with the cost function observables to create a classical quantity which generates the solution space. The optimizer traverses this solution space, moving from one solution candidate to another every iteration that corresponds to a slightly lower energy than the previous case.  

Hence, different choices and combinations of ansatz and cost functions generate different solution spaces with slightly varying solution candidates at every iteration. This further translates to some degree of differences in how the same optimizer will traverse the solution space for different ansatz and cost function combinations. Through a degree of hyperparameter optimization for different elements of VQE via simulator runs, and in addition to implementing the bitstring recovery mechanism (discussed in the next section) to correct bitstring errors, we observe a good improvement in performance, as discussed in Section \ref{sec:results}. 

The overall workflow can be broken down into stages as given by the Qiskit pattern framework \cite{qiskit_pattern}. The stages are known as $(i)$ Map, $(ii)$ Optimize, $(iii)$ Execute and $(iv)$ Post-process. The breaking down into stages allow for a quantum computing problem to be executed seamlessly by heterogeneous computing infrastructure. 


\subsubsection{\textbf{Recovery Mechanism for Bitstrings}} \label{sec:bit_rec}

A well-constructed VQE should ideally traverse the cost-function landscape and converge to a local or global minimum (or, more generally, a stationary point). However, particularly for large problems that require a large number of qubits, the classical optimizer often struggles to effectively navigate the high-dimensional parameter space.
    
In this problem, we observe that for graphs with higher number of nodes (18-24 nodes using 90-120 qubits), simple CVaR VQE with yield bitstring sequences that have anomalies and violations (refer to Section \ref{sec:result-analysis} and Fig. \ref{fig:errors} for the types of errors). To address these errors (node repetition, invalid nodes, and invalid edges), we use a technique similar to self-consistent configuration recovery employed in the SQD algorithm and other works \cite{skqd}. 

In VQE, the quality of intermediate solutions strongly impacts the optimization trajectory and eventual convergence. Low-quality outputs can cause the optimizer to deviate from the ideal path, increasing the risk of convergence to suboptimal solutions. Ensuring high-quality intermediate solutions is therefore essential for guiding the optimizer effectively through the solution space. Imperfect cost function design and inherent noise and errors from quantum hardware severely impact the quality of intermediate solutions. 

We define a node as anomalous if it is repeated or lies outside the valid range of nodes. Section \ref{sec:result-analysis} provides a detailed discussion of the different types of anomalies and errors observed. There exists a surjective mapping from the set of anomalous nodes to the set of nodes missing from the output sequence. At each iteration, bitstrings containing anomalies can be corrected by applying an onto mapping from the anomalous nodes to the missing nodes. 

For example, consider the 6-node sequence given below. Clearly, node $2$ is repeated, and node $3$ is missing. By replacing node $2$ in the sequence with node $3$, the sequence can be corrected.
\begin{itemize}
    \item Incorrect sequence
    $$ 0 \rightarrow 1 \rightarrow 2\rightarrow 5 \rightarrow 2 \rightarrow 4 $$

    \item Corrected sequence
    $$0 \rightarrow 1 \rightarrow 2\rightarrow 5 \rightarrow 3 \rightarrow 4$$
\end{itemize}

Once a bitstring sample is acquired from the hardware output, we process the bitstring before feeding it back to the optimizer. The algorithm to do that is as follows.

\begin{itemize}
    \item Input: A bitstring sample $b$ obtained from quantum hardware, encoding a sequence of nodes.

    \item Identify anomalies and missing nodes. Let $A$ be the set of anomalous nodes and $M$ be the set of missing nodes.

    \item Compute pairwise differences and create a difference matrix.
    For each $a \in A$ and $m \in M$, compute the Hamming distance
    $$ d(a,m) $$.

    \item Construct a greedy mapping based on the difference matrix. Let $\phi: A \rightarrow M$ be a mapping from anomalous to missing nodes. Select $(a^*, m^*)$.
    $$ (a^*, m^*)  = \argmin_{a \in A, m \in M} d(a,m)$$.

    \item Replace $a^*$ in the output sequence with the mapped node $ \phi(a^*) = m^*$.
\end{itemize}

\begin{figure*}[ht]
	\centering
	\includegraphics[width=1.0\linewidth]{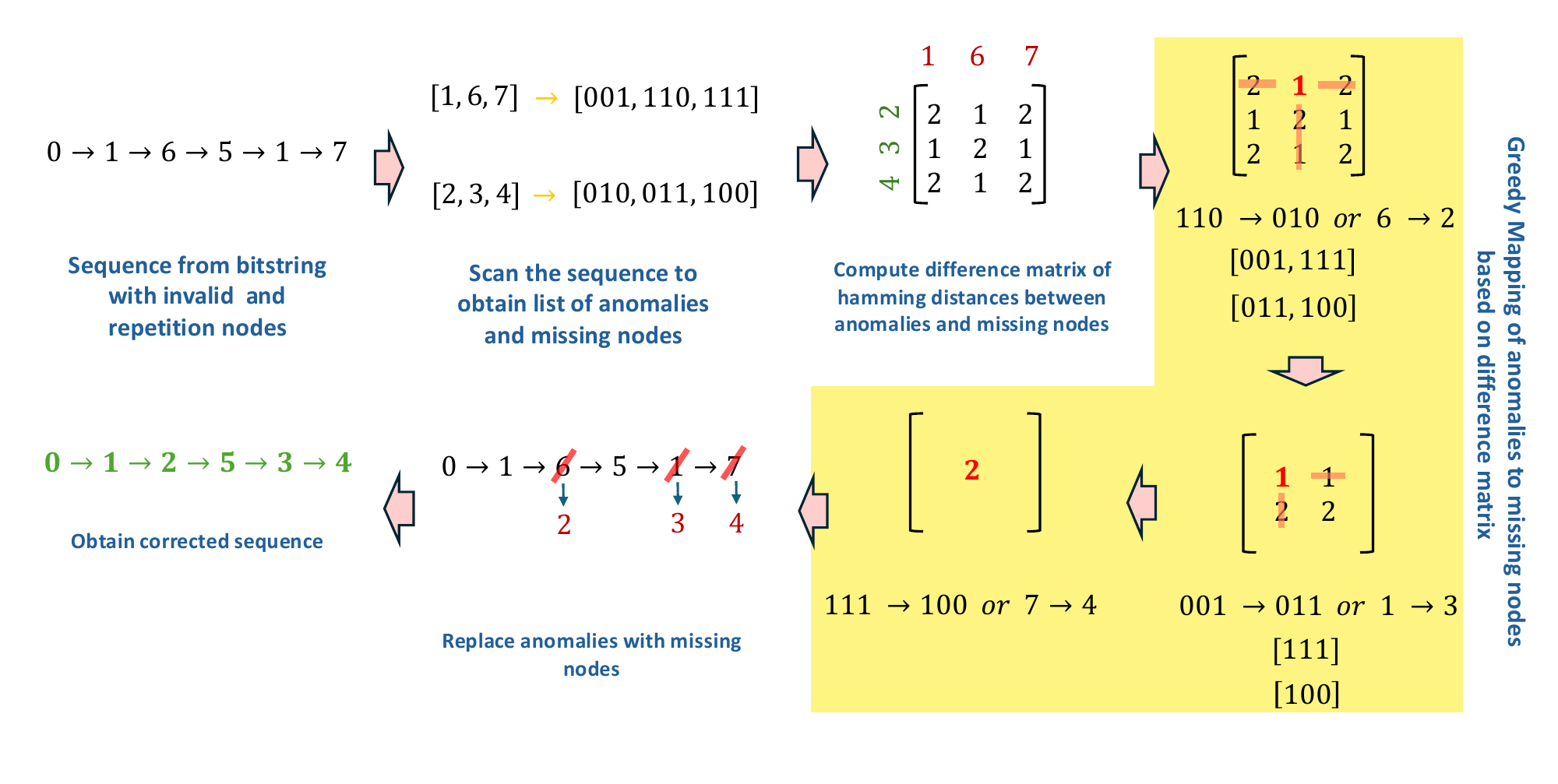}
	\caption{A demonstration of the self-consistent recovery mechanism for bitstrings applied to an example sequence containing repetitions and invalid nodes (anomalies)}
	\label{fig:config_recovery_steps}
\end{figure*}

As a result, we obtain a sequence that contains only valid nodes, in contrast to the original sequence that included invalid nodes. This improvement in quality of intermediate solutions leads to better optimizer trajectories (via improved expectation value estimation), ultimately resulting in faster convergence and convergence to higher-quality solutions. Fig. \ref{fig:config_recovery_steps} gives a demonstration of the recovery mechanism.

It is important to emphasize that this mechanism does not constitute error mitigation or configuration recovery in the conventional sense. The primary driver of convergence remains the optimizer, which iteratively updates the parameters under the guidance of the cost function. The role of the recovery mechanism is instead to support the optimization process by ensuring that, at each iteration, invalid configurations arising from anomalies do not interfere with the optimizer's traversal of the solution space and cause its trajectory to deviate away from regimes of convergence.

While the recovery mechanism facilitates improved traversal of the solution space by the optimizer, it does not guarantee convergence. Likewise, there may be instances in which the recovery step itself induces violations due to the nature of the mapping. This is acceptable, as the recovery mechanism is solely intended to keep the optimizer's trajectory within valid regions of the solution space, while the responsibility for achieving convergence is left entirely to the optimizer. The recovery mechanism offers two primary advantages. First, propagating high-quality, valid bitstrings across iterations improves overall solution quality. Second, it accelerates convergence by enabling the optimizer to reach the solution via high-quality intermediate solution candidates in fewer iterations.

\subsection{Contig Generation and Organism Identification}\label{method:contig_gen}
Once a valid path is obtained from the OLC graph using either quantum or classical solution methods, the next step is to convert this path into biologically meaningful information. Specifically, the ordered path of reads is used to construct a contig, representing a contiguous assembled segment of the genome. 

Contig generation begins with the first read, \( C^{(1)} = r_1 \), and proceeds sequentially by extending the contig along the ordered read path. For each subsequent read \( r_i \), the suffix of the current contig \( C^{(i-1)} \) and the prefix of \( r_i \) are extracted and aligned using the same local sequence alignment procedure. An alignment score \( \mathcal{S}_i \) is computed, and alignments satisfying \( \mathcal{S}_i \geq 0 \) are considered to represent valid overlaps.

\medskip
For a valid alignment, the effective overlap length \( \ell_i \) is defined as the number of aligned positions where neither sequence contains a gap:
\begin{equation}
\ell_i = \text{count of non-gap aligned positions between } C^{(i-1)} \text{ and } r_i.
\end{equation}

The contig is then extended by appending only the non-overlapping suffix of the incoming read, resulting in the updated contig:
\begin{equation}
C^{(i)} = C^{(i-1)} \;\text{concat}\!\left( r_i[\ell_i + 1, \text{end}] \right),   
\end{equation}
where concat denotes sequence concatenation and \( r_i[\ell_i + 1, \text{end}] \) represents the portion of \( r_i \) following the overlap region.

This procedure is repeated iteratively until all reads in the path are incorporated, yielding the final contig \( C^{(n)} \), which represents a partially assembled genome sequence. This approach enables the construction of high-confidence contigs and provides a partial assembled genome of the organism that can help in organism identification via genome database searches. BLASTn analyses were performed on the assembled contigs against the NCBI nucleotide database \cite{altschul1990basic}, restricted to available complete organism genome sequences. The closest matching organisms were identified based on query coverage and sequence identity, with the top-scoring alignments used to assign the most likely taxonomic origin of the \textit{de novo} assembled reads.

\clearpage
\subsection*{Supplementary Tables}
\makeatletter
\renewcommand{\fnum@table}{\textbf{\tablename\ \thetable}}
\makeatother

\begin{table}[h]
\centering
\caption{\label{table:adapters} Different types of adapters and their sequences.}
\begin{tabular}{ll}
\hline
Adapter type & Adapter Sequence \\
\hline
Illumina Universal Adapter      & AGATCGGAAGAG \\
Illumina Small RNA 3$'$ Adapter & TGGAATTCTCGG \\
Illumina Small RNA 5$'$ Adapter & GATCGTCGGACT \\
Nextera Transposase Sequence    & CTGTCTCTTATA \\
PolyA Adapter Sequence          & AAAAAAAAAA \\
PolyG Adapter Sequence          & GGGGGGGGGG \\
\hline
\end{tabular}
\end{table}

\begin{table}[h]
\centering
\caption{\label{tab:sra_datasets_Supp} Organisms and corresponding SRA read datasets used in this study \cite{katz2022sequence}}
\begin{tabular}{ll}
\hline
Organism & SRA FASTQ Dataset \\
\hline
\textit{Bacillus cereus} & SRR12045421 \\
\textit{Bacteroides intestinalis} & SRR13062400 \\
\textit{Avian influenza (HPAI) H5N1 virus - Brown\_Skua} & SRR32006043 \\
\textit{Dengue virus type 2 Bangladesh/NRICh-DS028/2023 }& SRR32137766 \\
\textit{Monkeypox virus Kenya isolate} & SRR32413059 \\
\textit{African swine fever virus isolate NAM P1/1995} & SRR27477754 \\
\hline
\end{tabular}

\vspace{0.5em}
\footnotesize
Small-genome organisms such as bacteria and viruses (5 - 80 Mb in size) are used to generate shorter graphs suitable for current quantum hardware availability.
\end{table}

\begin{table}[h]
\centering
\caption{Summary of read datasets, graph size, and organism identification from assembled contigs obtained using classical dynamic programming process.}
\label{tab:supp_classical_organism_identification}
\renewcommand{\arraystretch}{1.25}

\begin{tabularx}{\textwidth}{l c X c c}
\toprule
\textbf{Read dataset} &
\textbf{Graph nodes} &
\textbf{Organism identified from the assembled contig} &
\textbf{Query coverage (\%)} &
\textbf{Identity (\%)} \\
\midrule

SRR12045421 & 4 &
\textit{Bacillus thuringiensis} & 100 & 96.19 \\
 &  &
\textit{*Bacillus cereus} & 99 & 96.07 \\
\midrule

SRR32137766 & 6 &
\textit{Dengue virus type 2 isolate Phuket 2020} & 80 & 89.07 \\
 &  &
\textit{*hDenV2/Bangladesh/NRICh-DS028/2023} & 79 & 88.62 \\
\midrule

SRR13062400 & 7 &
\textit{*Bacteroides intestinalis} & 95 & 93.66 \\
\midrule

SRR32413059 & 9 &
\textit{Monkeypox virus isolate USA\_CA0103} & 100 & 96.5 \\
\midrule

SRR32006043 & 18 &
\textit{Influenza A (Antarctica\_Tern/H5N1)} & 83 & 91.69 \\
 &  &
\textit{*Influenza A (Brown\_Skua/H5N1)} & 83 & 91.69 \\
\midrule

SRR27477754 & 21 &
\textit{*African swine fever virus isolate NAM/P1/1995} & 56 & 91.81 \\
\midrule

SRR27477754 & 24 &
\textit{*African swine fever virus isolate NAM/P1/1995} & 78 & 94.26 \\
\midrule

\bottomrule
\end{tabularx}
\vspace{0.5em}
\footnotesize
* Represents the actual organism and the strain, the SRA dataset belongs to.
\end{table}

\FloatBarrier
\renewcommand{\refname}{References}
\renewcommand{\bibname}{References}

\putbib
\end{bibunit}

\end{document}